\documentclass[twocolumn,showpacs,aps,prl,superscriptaddress]{revtex4}

\usepackage{graphicx}
\usepackage{dcolumn}
\usepackage{amsmath}
\usepackage{epsfig}

\RequirePackage{xspace}

\usepackage{relsize}
\def\babar{\mbox{\slshape B\kern-0.1em{\smaller A}\kern-0.1em
    B\kern-0.1em{\smaller A\kern-0.2em R}}}

\def\epem       {\ensuremath{e^+e^-}\xspace}

\def\q     {\ensuremath{q}\xspace}

\def\qqbar {\ensuremath{q\overline q}\xspace}

\def\d     {\ensuremath{d}\xspace}

\def\s     {\ensuremath{s}\xspace}

\def\ssbar {\ensuremath{s\overline s}\xspace}

\def\ccbar {\ensuremath{c\overline c}\xspace}
\def\b     {\ensuremath{b}\xspace}

\def\pip   {\ensuremath{\pi^+}\xspace}
\def\pim   {\ensuremath{\pi^-}\xspace}

\def\kaon  {\ensuremath{K}\xspace}
\def\Kbar  {\kern 0.2em\overline{\kern -0.2em K}{}\xspace}

\def\Kz    {\ensuremath{K^0}\xspace}
\def\Kzb   {\ensuremath{\Kbar^0}\xspace}
\def\KzKzb {\ensuremath{\Kz \kern -0.16em \Kzb}\xspace}
\def\Kp    {\ensuremath{K^+}\xspace}
\def\Km    {\ensuremath{K^-}\xspace}

\def\KpKm  {\ensuremath{\Kp \kern -0.16em \Km}\xspace}
\def\KS    {\ensuremath{K^0_{\scriptscriptstyle S}}\xspace}

\def\Kstarzb {\ensuremath{\Kbar^{*0}}\xspace}

\def\Dbar    {\kern 0.2em\overline{\kern -0.2em D}{}\xspace}

\def\Dz      {\ensuremath{D^0}\xspace}
\def\Dzb     {\ensuremath{\Dbar^0}\xspace}
\def\DzDzb   {\ensuremath{\Dz {\kern -0.16em \Dzb}}\xspace}
\def\Dp      {\ensuremath{D^+}\xspace}
\def\Dm      {\ensuremath{D^-}\xspace}

\def\DpDm    {\ensuremath{\Dp {\kern -0.16em \Dm}}\xspace}

\def\B       {\ensuremath{B}\xspace}
\def\Bbar    {\kern 0.18em\overline{\kern -0.18em B}{}\xspace}

\def\BB      {\ensuremath{B\Bbar}\xspace} 
\def\Bz      {\ensuremath{B^0}\xspace}
\def\Bzb     {\ensuremath{\Bbar^0}\xspace}
\def\BzBzb   {\ensuremath{\Bz {\kern -0.16em \Bzb}}\xspace}
\def\Bu      {\ensuremath{B^+}\xspace}
\def\Bub     {\ensuremath{B^-}\xspace}
\def\Bp      {\ensuremath{\Bu}\xspace}

\def\BpBm    {\ensuremath{\Bu {\kern -0.16em \Bub}}\xspace}

\def\BorBbar    {\kern 0.18em\optbar{\kern -0.18em B}{}\xspace}
\def\DorDbar    {\kern 0.18em\optbar{\kern -0.18em D}{}\xspace}
\def\KorKbar    {\kern 0.18em\optbar{\kern -0.18em K}{}\xspace}

\def\jpsi     {\ensuremath{{J\mskip -3mu/\mskip -2mu\psi\mskip 2mu}}\xspace}
\def\psitwos  {\ensuremath{\psi{(2S)}}\xspace}

\mathchardef\Upsilon="7107
\def\Y#1S{\ensuremath{\Upsilon{(#1S)}}\xspace}

\def\FourS {\Y4S}

\mathchardef\Deltares="7101
\mathchardef\Xi="7104
\mathchardef\Lambda="7103
\mathchardef\Sigma="7106
\mathchardef\Omega="710A

\def\Deltabar{\kern 0.25em\overline{\kern -0.25em \Deltares}{}\xspace}
\def\Lbar{\kern 0.2em\overline{\kern -0.2em\Lambda\kern 0.05em}\kern-0.05em{}\xspace}
\def\Sigbar{\kern 0.2em\overline{\kern -0.2em \Sigma}{}\xspace}
\def\Xibar{\kern 0.2em\overline{\kern -0.2em \Xi}{}\xspace}
\def\Obar{\kern 0.2em\overline{\kern -0.2em \Omega}{}\xspace}
\def\Nbar{\kern 0.2em\overline{\kern -0.2em N}{}\xspace}
\def\Xb{\kern 0.2em\overline{\kern -0.2em X}{}\xspace}

\def\mes        {\mbox{$m_{\rm ES}$}\xspace}

\def\DeltaE     {\mbox{$\Delta E$}\xspace}

\newcommand{\tev}{\ensuremath{\mathrm{\,Te\kern -0.1em V}}\xspace}
\newcommand{\gev}{\ensuremath{\mathrm{\,Ge\kern -0.1em V}}\xspace}
\newcommand{\mev}{\ensuremath{\mathrm{\,Me\kern -0.1em V}}\xspace}
\newcommand{\kev}{\ensuremath{\mathrm{\,ke\kern -0.1em V}}\xspace}
\newcommand{\ev}{\ensuremath{\mathrm{\,e\kern -0.1em V}}\xspace}
\newcommand{\gevc}{\ensuremath{{\mathrm{\,Ge\kern -0.1em V\!/}c}}\xspace}
\newcommand{\mevc}{\ensuremath{{\mathrm{\,Me\kern -0.1em V\!/}c}}\xspace}
\newcommand{\gevcc}{\ensuremath{{\mathrm{\,Ge\kern -0.1em V\!/}c^2}}\xspace}
\newcommand{\mevcc}{\ensuremath{{\mathrm{\,Me\kern -0.1em V\!/}c^2}}\xspace}

\def\invfb   {\ensuremath{\mbox{\,fb}^{-1}}\xspace}

\def\mus  {\ensuremath{\rm \,\mus}\xspace}

\def\mus        {\ensuremath{\,\mu{\rm s}}\xspace}

\def\to                 {\ensuremath{\rightarrow}\xspace}

\def\pep2{PEP-II}

\newcommand{\dedx}{\ensuremath{\mathrm{d}\hspace{-0.1em}E/\mathrm{d}x}\xspace}

\def\gsim{{~\raise.15em\hbox{$>$}\kern-.85em
          \lower.35em\hbox{$\sim$}~}\xspace}
\def\lsim{{~\raise.15em\hbox{$<$}\kern-.85em
          \lower.35em\hbox{$\sim$}~}\xspace}

\def\CP                {\ensuremath{C\!P}\xspace}

\xspace

\def\jetset74   {\mbox{\tt Jetset \hspace{-0.5em}7.\hspace{-0.2em}4}\xspace}

\renewcommand{\eqref}[1]{Eq.~(\ref{eq:#1})}

\newcommand{\splot}    {\mbox{$_s{\cal P}lot$}\xspace}

\newcommand{\onreslumi}  {\mbox{347.5\invfb}}
\newcommand{\offreslumi} {\mbox{36.6\invfb}}
\newcommand{\bbpairs}    {\mbox{$(383.2\pm4.2)\times10^{6}$}}

\def\ncand    {\ensuremath{16143}}
\def\nsig     {\ensuremath{429\pm 43}}
\def\nsigma   {\ensuremath{12.6\,\sigma}}
\def\nsigmatot{\ensuremath{9.6\,\sigma}}
\def\nqqbar   {\ensuremath{14850\pm 129}}
\def\kkpiBF   {\ensuremath{\left(5.0\pm 0.5\pm 0.5\right)\times 10^{-6}}}
\def\kkpiBFwe {\ensuremath{[5.0\pm 0.5 ({\rm stat})\pm 0.5 ({\rm syst})]\times 10^{-6}}}
\def\kkpiAcp  {\ensuremath{0.00\pm 0.10\pm 0.03}}
\def\nnout    {\ensuremath{NN_{\rm out}}}
\def\nnprim   {\ensuremath{NN^\prime_{\rm out}}}
\def\nnprime  {\ensuremath{NN^{\prime\,\,\,i}_{\rm out}}} 

\newcommand{\BABARPubYear}    {07}
\newcommand{\BABARPubNumber}  {041}
\newcommand{\SLACPubNumber} {12702}
\newcommand{\LANLNumber} {arXiv:0708.0376 [hep-ex]}

\def\figurebox#1#2#3{
    \def\arg{#3}
    \ifx\arg\empty
    {\hfill\vbox{\hsize#2\hrule\hbox to #2{\vrule\hfill\vbox to #1{\hsize#2\vfill}\vrule}\hrule}\hfill}
    \else
    {\hfill\epsfbox{#3}\hfill}
    \fi}

\begin{document}

\preprint{\babar-PUB-\BABARPubYear/\BABARPubNumber} 
\preprint{SLAC-PUB-\SLACPubNumber} 

\begin{flushleft}
\babar-PUB-\BABARPubYear/\BABARPubNumber\\
SLAC-PUB-\SLACPubNumber\\
\LANLNumber\\[10mm]
\end{flushleft}

\title{
  {\large 
    \bf Observation of the Decay {\boldmath $\Bu\to\Kp\Km\pip$}
  }
}

\author{B.~Aubert}
\author{M.~Bona}
\author{D.~Boutigny}
\author{Y.~Karyotakis}
\author{J.~P.~Lees}
\author{V.~Poireau}
\author{X.~Prudent}
\author{V.~Tisserand}
\author{A.~Zghiche}
\affiliation{Laboratoire de Physique des Particules, IN2P3/CNRS et Universit\'e de Savoie, F-74941 Annecy-Le-Vieux, France }
\author{J.~Garra~Tico}
\author{E.~Grauges}
\affiliation{Universitat de Barcelona, Facultat de Fisica, Departament ECM, E-08028 Barcelona, Spain }
% ...
\collaboration{\babar\ Collaboration}
\noaffiliation

\date{\today}

\begin{abstract}
We report the observation of charmless hadronic decays
of charged \B\ mesons to the final state $\Kp\Km\pip$. 
Using a data sample of \onreslumi\ collected at the \FourS
resonance with the \babar\ detector,
we observe \nsig\ signal events with a significance of \nsigmatot. 
We measure the inclusive branching fraction
${\cal B}(\Bu\to\Kp\Km\pip)=\kkpiBFwe$.
Inspection of the Dalitz plot of signal candidates 
shows a broad structure peaking
near 1.5\,\gevcc\ in the $\Kp\Km$ invariant mass distribution.
We find the direct \CP\ asymmetry to be consistent with zero.
\end{abstract}

\pacs{13.25.Hw, 12.15.Hh, 11.30.Er}

\maketitle

\B\ meson decays to final states with even numbers of strange
quarks or antiquarks are suppressed in the Standard Model.
Such decays may proceed by the \b\to\d\ loop (penguin) transition,
or by other processes followed by \ssbar\ production. Hadronic
\b\to\d\ penguin transitions have recently been
observed~\cite{Aubert:2006gm,Abe:2006xs}, while examples of
\ssbar\ production have been seen in various
\B\ decays~\cite{Drutskoy:2002ib,Krokovny:2002pe,Aubert:2006hu}.
Furthermore, Dalitz plot (DP) analyses of 
$\Bu\to\Kp\Kp\Km$~\cite{Garmash:2004wa,Aubert:2006nu} and
$\Bz\to\Kp\Km\Kz$~\cite{babar:kkks} have seen anomalous excesses
of events at low $\Kp\Km$ invariant masses, the origin of which
has aroused considerable interest among theorists~\cite{Minkowski:2004xf,
Furman:2005xp,Cheng:2005nb,Wang:2006ri,Cheng:2007si},
as it is of great importance in the understanding of low energy spectroscopy~\cite{Klempt:2007cp}.
Understanding the production mechanism of charmless $\B$ decays 
to such multibody final states is therefore a priority.

The decay $\Bp\to\Kp\Km\pip$ and charmless quasi-two-body \B\ decays
resulting in this final state have not been previously observed.
The current experimental upper limits are 
${\cal B}(\Bu\to\Kp\Km\pip)<6.3\times 10^{-6}$~\cite{Aubert:2003xz}, 
${\cal B}(\Bu\to\phi\pip)<2.4\times 10^{-7}$~\cite{Aubert:2006nn}
and
${\cal B}(\Bu\to\Kp\Kstarzb(892))<1.1\times 10^{-6}$~\cite{Aubert:2007ua},
all at 90\,\% confidence level.
Such decays play an important role in analyses based on flavor SU(3)
that can limit the allowed values of the deviation of
$\sin(2\beta^{\rm eff})$ measured in hadronic \b\to\s\ penguin
modes to the reference value obtained in $\b\to\ccbar\s$ transitions
such as $\Bz\to\jpsi\KS$~\cite{theory:su3}.
Various theoretical predictions give 
${\cal B}(\Bu\to\phi\pip) \lsim 
{\cal O}(10^{-8})$~\cite{Chiang:2003pm,Du:2002up,Beneke:2003zv,
  Bar-Shalom:2002sv,Giri:2004wn} and 
${\cal B}(\Bu\to\Kp\Kstarzb(892)) \lsim 
{\cal O}(10^{-6})$~\cite{Chiang:2003pm,Du:2002up,Beneke:2003zv,
  Guo:2006uq,Ali:1998eb}.
A recent phenomenological analysis gives a lower bound of 
${\cal B}(\Bu\to\Kp\Kstarzb(892)) \gsim 
0.7 \times 10^{-6}$~\cite{theory:kstark_bound}.

We report herein the results of a search for
the charmless hadronic decay $\Bu\to\Kp\Km\pip$~\cite{cc}. 
The data used in this analysis, collected
at the \pep2\ asymmetric energy \epem collider~\cite{pep2}, 
consist of an integrated luminosity of \onreslumi\ 
recorded at the \FourS\ resonance. 
In addition, \offreslumi\ of data were collected 40\,\mev\ below 
the resonance. These samples are referred to as
on-resonance and off-resonance data, respectively.
The on-resonance data sample contains \bbpairs\ \BB\ pairs~\cite{Aubert:2002hc}.

The \babar\ detector is described in detail elsewhere \cite{Aubert:2001tu}.
Charged particles are detected and their momenta measured
with a five-layer silicon vertex tracker (SVT) and a 40-layer
drift chamber (DCH) inside a 1.5\,T solenoidal magnet. Surrounding
the DCH is a detector of internally reflected Cherenkov radiation
(DIRC), designed for charged particle identification (PID). 
Energy deposited by electrons and photons
is measured by a CsI(Tl) crystal electromagnetic calorimeter.

We select $\Bu\to\Kp\Km\pip$ candidates by combining two charged kaon
candidates of opposite sign with one charged pion candidate.
Each track is required to have at least 12 hits in the DCH, 
to have a minimum transverse momentum of 100\,\mevc, 
and to be consistent with having originated from the interaction region. 
Identification of charged pions and kaons is accomplished using
energy loss (\dedx) information from the SVT and DCH, 
and the Cherenkov angle and number of photons measured 
by the DIRC for tracks with momenta above 700\,\mevc. 
We distinguish kaons from pions by applying criteria to the
product of the likelihood ratios determined from these individual
measurements. The efficiency for kaon selection is approximately
80\,\% including geometrical acceptance, while the probability
of misidentification of pions as kaons is below 5\,\% up to a laboratory
momentum of 4\,\gevc.

Continuum $\epem\to\qqbar \ (\q=u,d,s,c)$ events are the dominant
background. 
To discriminate this type of event from signal,
we use a neural network~\cite{ref:TMVA} that combines five variables: 
the ratio of the second order momentum-weighted Legendre polynomial 
moment to that of the zeroth order; 
the absolute value of the cosine of the angle between the \B
direction and the beam ($z$) axis;
the magnitude of the cosine of the angle between the \B\ thrust axis 
and the $z$ axis [all quantities calculated in the center-of-mass (c.m.) frame];
the product of the \B\ candidate's charge and the flavor of the recoiling \B
as reported by a multivariate tagging algorithm~\cite{Aubert:2002ic};
and the boost-corrected proper time difference
between the decays of the two \B\ mesons divided by its variance.

In addition to the neural network output (\nnout),
we distinguish signal from background events using two kinematic variables: 
the difference \DeltaE\ between the c.m. energy
of the \B\ candidate and $\sqrt{s}/2$, 
and the beam-energy substituted mass
$\mes=\sqrt{s/4-{\bf p}^2_\B}$,
where $\sqrt{s}$ is the total c.m. energy
and ${\bf p}_\B$ is the momentum of the candidate \B\ meson in the c.m. frame.
We select signal candidates that satisfy $\nnout>0.29$,
$5.272<\mes<5.286\gevcc$ and $|\DeltaE|<0.075\gev$.

Another potentially large source of background arises from \B\ decays 
containing charm mesons and charmonia.
We veto \B\ candidates with $\Kp\Km$ invariant mass 
within $-3\,\sigma$,\, $+5\,\sigma$ of the nominal \Dz\ mass, 
or with invariant mass of the $\Km\pip$ system within
$\pm4\,\sigma$ of the mass of the \jpsi\ or $\psitwos$~\cite{Yao:2006px}. 
Here, $\sigma$ is 25\,\mevcc\ for \Dz, and 21\,\mevcc\ for \jpsi\ and \psitwos.
The asymmetric \Dz\ veto is chosen to remove
backgrounds resulting from $\pi\to\kaon$ misidentification.
Charmonium contributions arise mainly from the leptonic
decays of \jpsi\ and \psitwos, when one lepton is misidentified
as a pion and the other as a kaon.

The efficiency for signal events to pass the selection criteria is 22.1\%,
determined with a Monte Carlo (MC) simulation in which events uniformly
populate the Dalitz plot.
The only selection requirements that exhibit any strong dependency
on the DP position are the track preselection 
(due to the reduced acceptance of low momentum tracks),
and charm and charmonia vetoes.
The average number of \B\ candidates found per selected event is 1.12.
In events with multiple candidates
we choose the one with the highest probability of a fit of the three tracks
to a common vertex. 
In about 1\,\% of signal events the \B\ candidate is misreconstructed
due to one track being replaced with a track from the rest of the event.
Such events are considered as a part of the signal component.

We study possible residual backgrounds from \BB\ events using MC simulations.
We find that these can be conveniently divided into three categories,
each having similar shapes in \DeltaE\ and \mes. 
The first two ($\BB_1$ and $\BB_2$) 
are dominated by specific decays, 
$\Bu\to\Kp\pip\pim$ and $\Bu\to\Kp\Kp\Km$ respectively.
The third category ($\BB_3$) contains the remainder of the \BB\ background,
and is mainly combinatoric in nature.
Based on our MC studies, the total number of \BB\ pairs in our data sample, 
and the branching fractions listed by~\cite{Yao:2006px,Barberio:2007cr},
we expect $69$, $255$, and $528$ events 
from the three \BB\ background categories, respectively.

In order to obtain the $\Bu\to\Kp\Km\pip$ signal yield,
we perform an unbinned extended maximum likelihood fit to the
candidate events using three
input variables: \mes, \DeltaE\ and $\nnprim=1-\arccos(2\nnout-1)$.
The \nnprim\ variable is designed to allow simpler modeling 
of the strongly peaking structures near zero
for continuum background and near one for signal. 
For each event category $j$ 
(signal, continuum background, or one of the three \BB\ background components),
we define a probability density function (PDF):
\begin{equation}
  \label{PDF-exp}
  {\cal P}^i_j \equiv
  {\cal P}_j(\mes^i) {\cal P}_j(\DeltaE^i) {\cal P}_j(\nnprime),
\end{equation}
where $i$ denotes the event index. 
This form of the PDF is found to be valid 
since correlations among the input variables are small.
The extended likelihood function is:
\begin{equation}
  \label{eq:extML-Eq}
  {\cal L} = 
  \prod_{k}\exp\left(-n_k\right)
  \prod_{i}\left[ \sum_{j}n_j{\cal P}^i_j \right],
\end{equation}
where $n_j$ is the yield belonging to the event category $j$.

The signal \mes\ and \DeltaE\ shapes are parametrized with the sum
of a Gaussian and a Crystal Ball function~\cite{ref:xtalBall} and the
sum of two Gaussians, respectively. 
We fix the shape parameters to the values obtained from the 
$\Bu\to\Kp\Km\pip$ phase space MC sample. 
The continuum background \mes\ shape is described by the function
$x\sqrt{1-x^2}\exp\left[-\xi(1-x^2)\right]$, with $x\equiv 2\mes/\sqrt{s}$
and $\xi$ a free parameter~\cite{Albrecht:1990am},
while the continuum \DeltaE\ shape is modeled with a linear function.
The \mes\ PDFs for two of the \BB\ background components are 
a Gaussian ($\BB_1$) and the sum of two Gaussians ($\BB_2$)
while those for the \DeltaE\ PDFs are 
the sum of a Gaussian and a linear function ($\BB_1$ and $\BB_2$).
The $\BB_3$ background category has the same functional forms 
as continuum in both \mes\ and \DeltaE,
and discrimination between these categories is provided only by \nnprim.
We use one-dimensional histograms to describe all $\nnprim$ distributions. 
These are obtained from MC samples for signal and the 
\BB\ background categories, and, for the continuum background,
from a combination of off-resonance data and 
on-resonance data in a continuum-dominated sideband of \mes\ and \DeltaE.

The free parameters of our fit are the signal and continuum yields, 
together with the $\xi$ parameter of the continuum \mes\ shape
and the slope of the continuum \DeltaE\ shape. 
All shape parameters and yields of the three \BB\ background categories are
fixed according to the MC expectations. 
All \nnprim\ shapes are fixed. 

We test the fitting procedure by applying it to ensembles
of simulated experiments where events are drawn from the PDF shapes
as described above for all five categories of events. We repeat the
exercise with \qqbar\ events alone drawn from the PDF into which we
embed signal and \BB\ background events
randomly extracted from the MC samples. 
We find negligible bias on the fitted signal yield in either case.

\begin{figure}[!htb]
\center
\includegraphics[width=.4942\columnwidth]{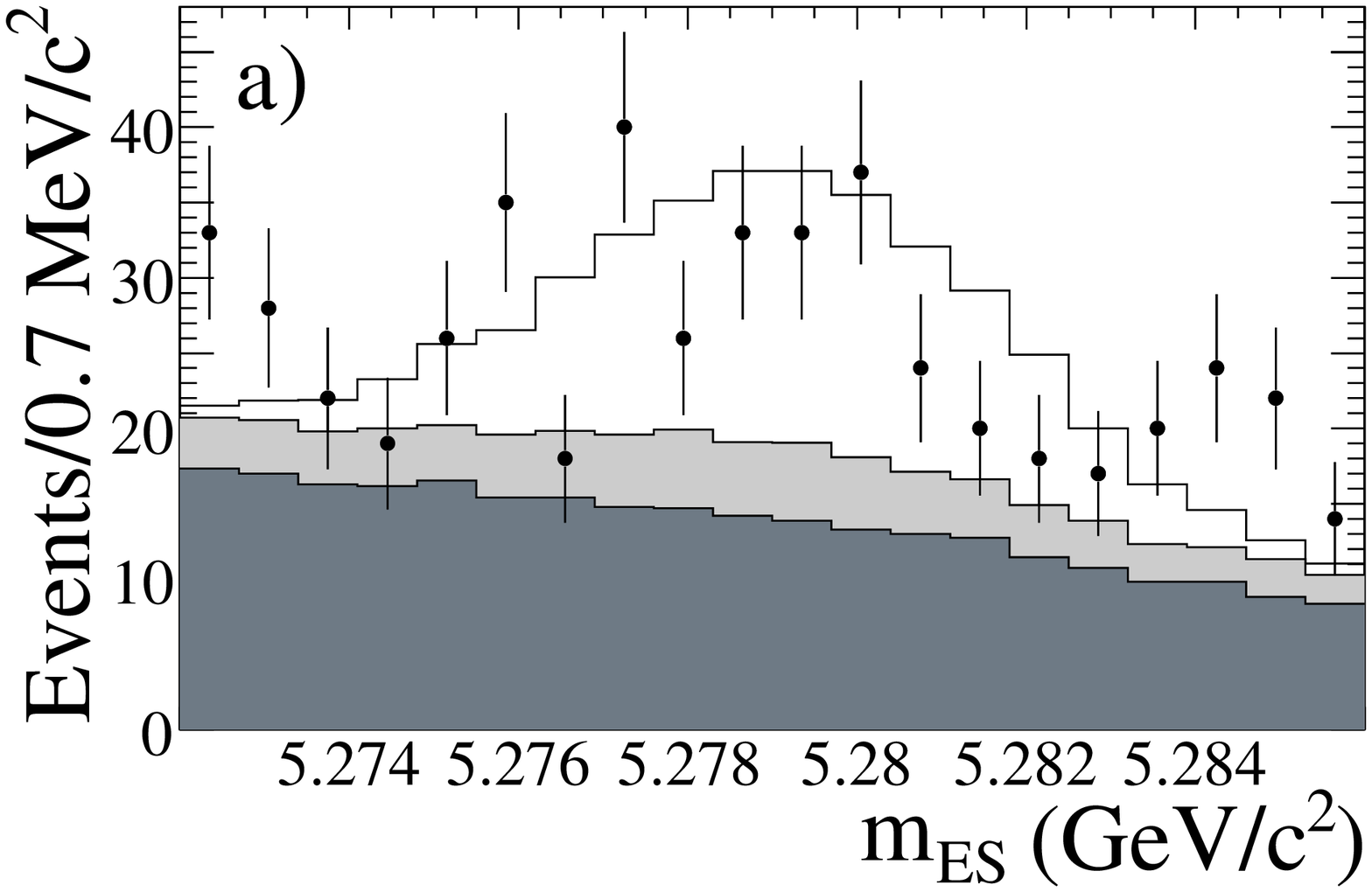}
\includegraphics[width=.4942\columnwidth]{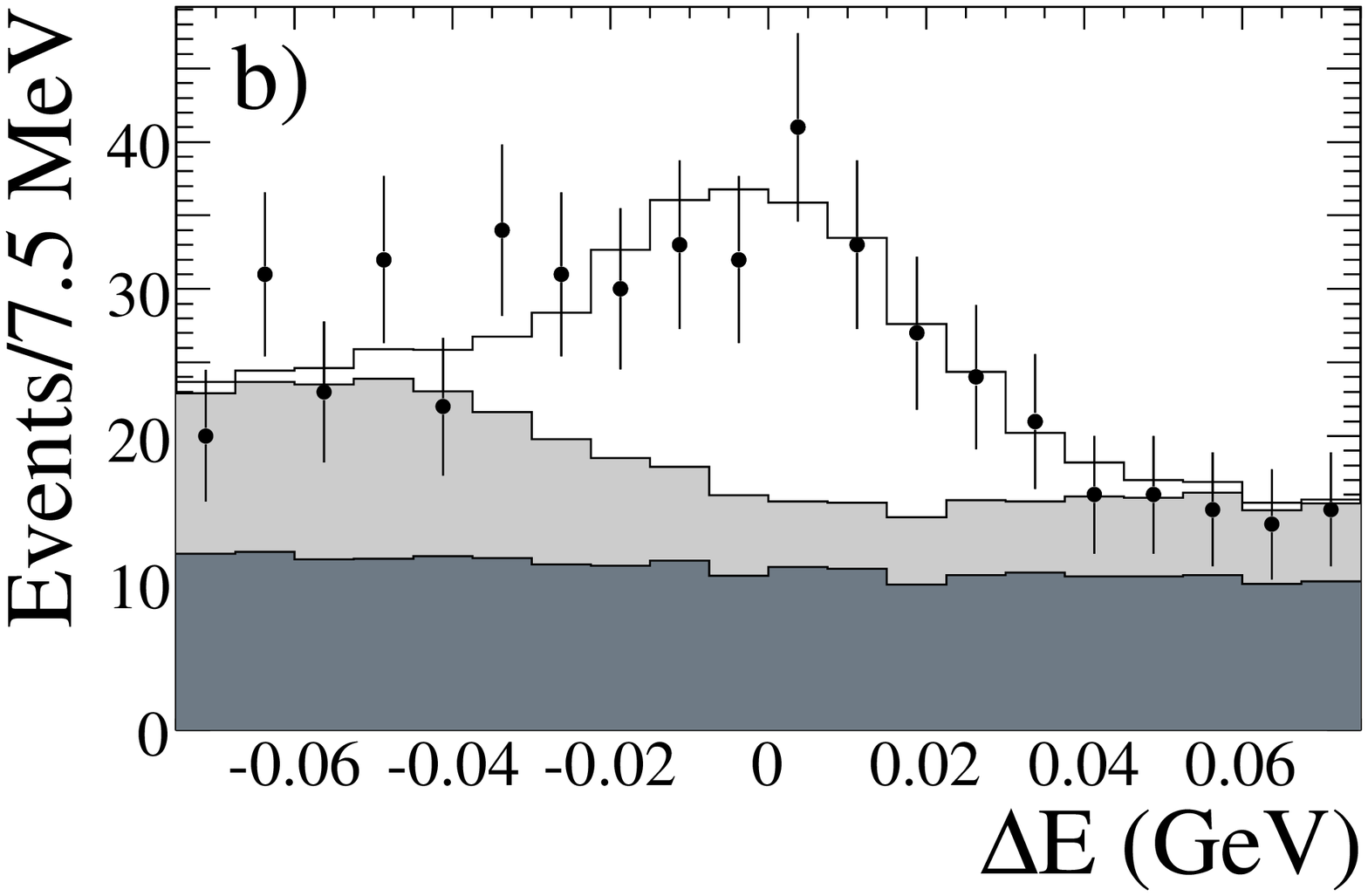}
\caption{
  Projections of candidate events onto
  a) \mes\ and b) \DeltaE\ following a signal enhancing cut 
  on the likelihood ratio calculated without the plotted variable. 
  Points show the data, dark filled histograms
  show the \qqbar\ background and light filled histograms show
  the \BB\ background component.
}
\label{fig:signal-splots}
\end{figure}

Using the fit described above to the \ncand\ candidate events,
we find \nsig\ signal events and \nqqbar\ \qqbar\ background events.
The results of the fit are shown in Fig.~\ref{fig:signal-splots}.
Both the \mes\ and \DeltaE\ distributions show clear signal peaks.
The statistical significance of the signal yield,
calculated from the change in negative log likelihood with signal yield
floated compared to that with signal yield fixed to zero is \nsigma.

We obtain the inclusive branching fraction of $\Bu\to\Kp\Km\pip$ using
the result of the fit to calculate 
signal probabilities for each candidate event~\cite{Pivk:2004ty}.
These are divided by event-by-event efficiencies,
that take the DP position dependence into account,
and summed to obtain an efficiency-corrected signal yield.
We further correct for the effect of the charm and charmonia vetoes,
and divide by the total number of $\BB$ events in the data sample.
The result is ${\cal B}\left(\Bu\to\Kp\Km\pip\right) = \kkpiBF$,
where the first error is statistical and the second is systematic.
The systematic error arises due to uncertainties in the PDF shapes (2.8\,\%)
including possible data-MC differences in the signal PDF shapes
studied using a control sample of $\Bu\to\Dzb\pip$, $\Dzb\to\Kp\pim$;
potential fit biases, dominated by the change in the result when
the yields of the \BB\ background components are floated (6.1\,\%);
uncertainties in the efficiency, due to tracking (2.4\,\%) and PID (4.2\,\%);
uncertainty in the correction due to vetoes, arising from the nonuniform
DP structure of the signal, and estimated from MC simulations
with different resonant contributions (6.1\,\%);
and the error in the number of \BB\ pairs (1.1\,\%).
The significance of the signal including systematic uncertainties is
found to be \nsigmatot\ from the change in negative log likelihood
with and without the signal component, while varying those sources
of uncertainty that affect the signal yield (PDF shapes and yields
of \BB\ background components).

We also extract the direct \CP\ asymmetry in the inclusive
signal yield by separately fitting $\B^-$ and $\B^+$ samples. 
The asymmetry is obtained using ${\cal A}_{\CP}=\frac{N^--N^+}{N^-+N^+}$ where
$N^-$ ($N^+$) is the fitted signal yield in the $B^-$ ($B^+$) sample,
corrected for efficiency and veto requirements. 
We find ${\cal A}_{\CP} = \kkpiAcp$, where the first error
is statistical and the second systematic, 
including uncertainties in the $\BB$ background estimation (0.02)
and possible detector asymmetry (0.02).
Other possible sources of systematic error are found to be negligible.

\begin{figure}[!htb]
\center
 \includegraphics[width=.97\columnwidth]{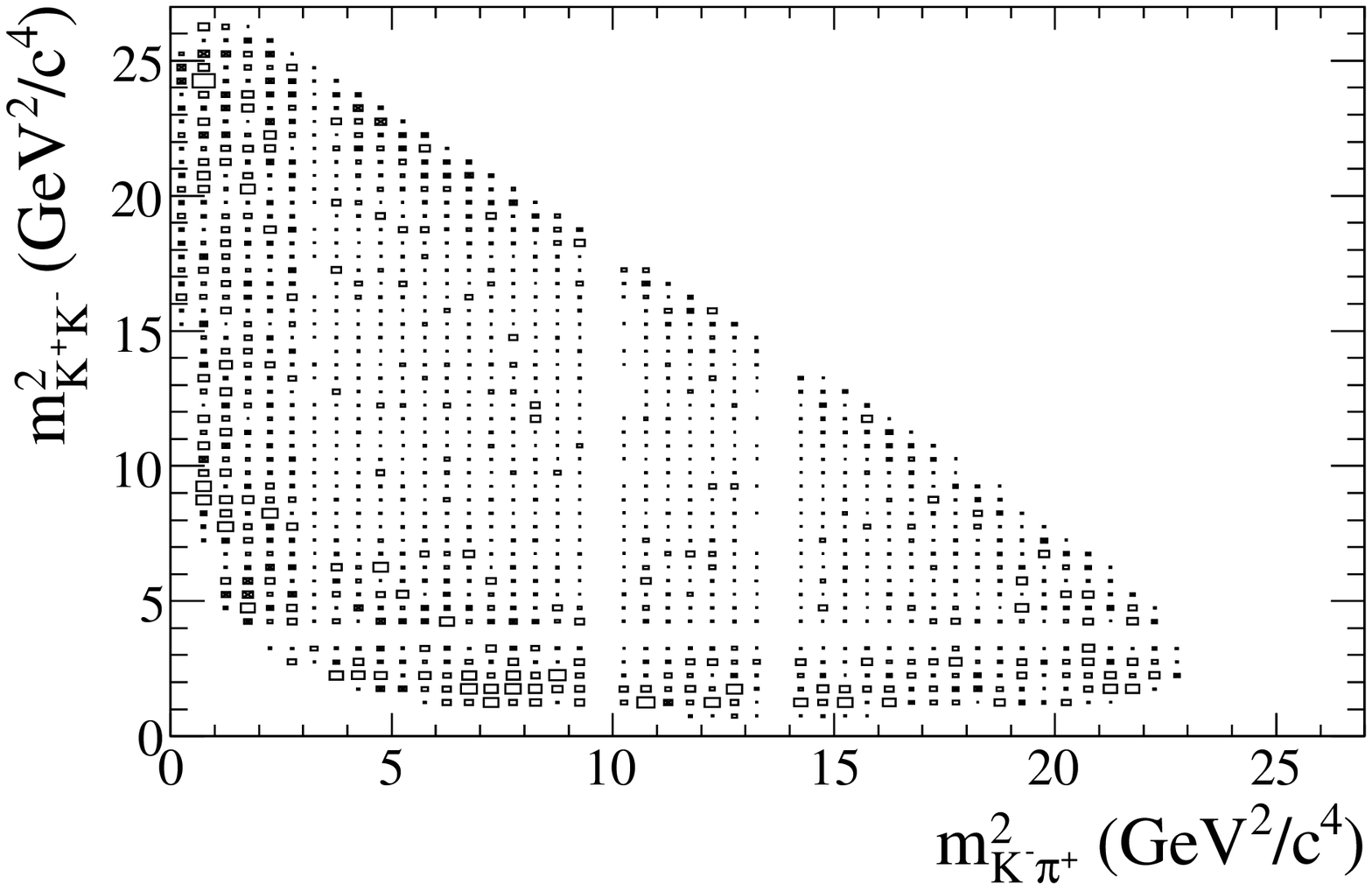}
\caption{
  Efficiency-corrected Dalitz plot distribution of $\Bu\to\Kp\Km\pip$ decays,
  obtained with the \splot\ technique~\cite{Pivk:2004ty}. 
  Empty regions correspond to charm and charmonia vetoes while the area
  of the boxes is proportional to the number of events in that bin.
}
\label{fig:signal-dalitz-splot}
\end{figure}

\begin{figure}[!htb]
\center
\includegraphics[width=.4942\columnwidth]{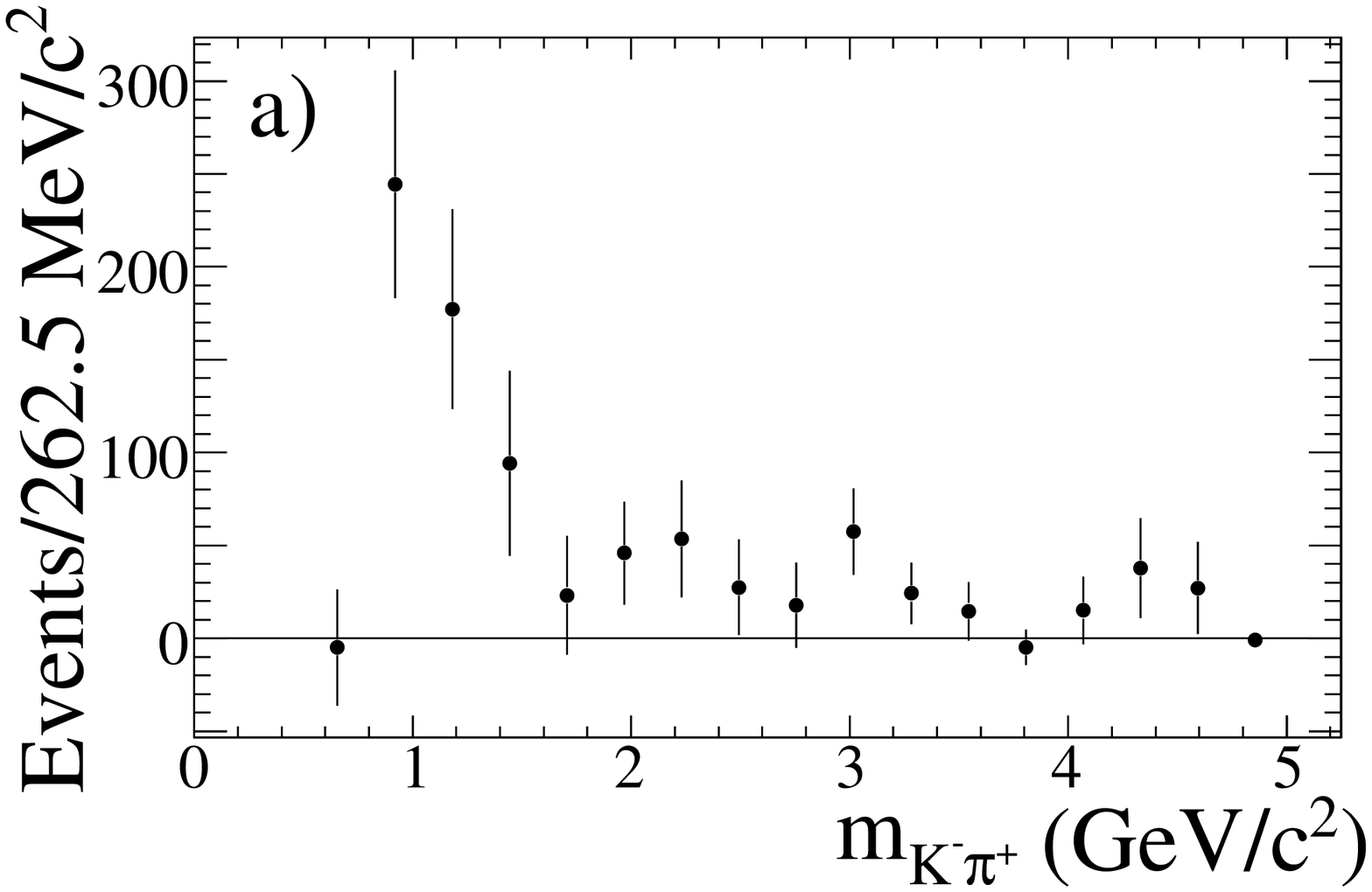}
\includegraphics[width=.4942\columnwidth]{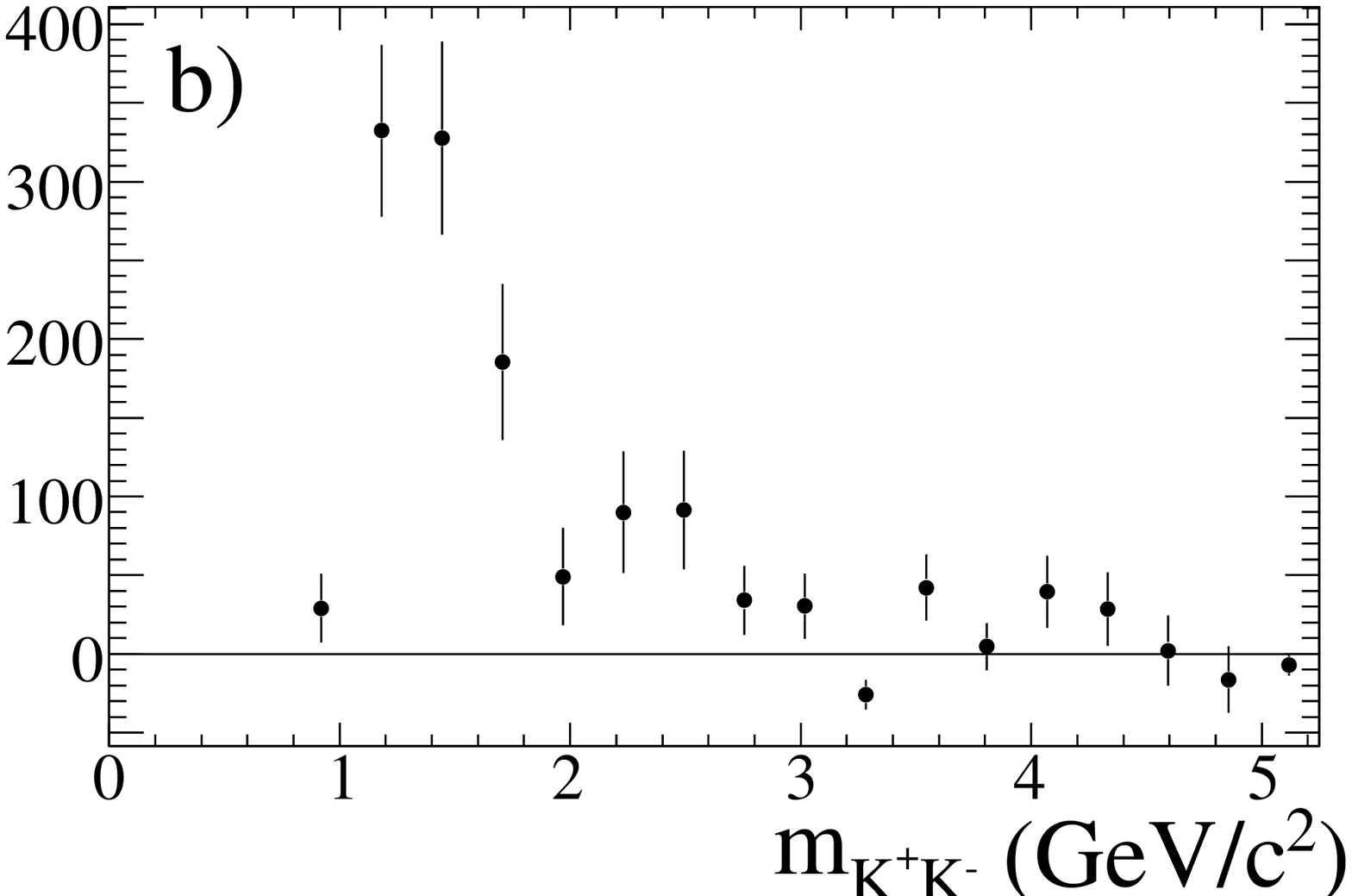}
\caption{
  Efficiency-corrected distributions of the $\Bu\to\Kp\Km\pip$
  signal candidates: a) $m_{\Km\pip}$ with $m_{\Kp\Km}>2.0\gevcc$ and
  b) $m_{\Kp\Km}$ with $m_{\Km\pip}>1.5\gevcc$.
  These projection plots are obtained with the \splot\ technique~\cite{Pivk:2004ty}.
}
\label{fig:fXplots}
\end{figure}

The efficiency-corrected Dalitz plot for signal decays,
obtained using event-by-event signal probabilities,
is shown in Fig.~\ref{fig:signal-dalitz-splot}. 
We have checked that this technique correctly reconstructs the 
signal DP distribution using MC simulations
in which the $\Bu\to\Kp\Km\pip$ events contain different structures.
In the data, we see an excess of events at low $\Km\pip$ invariant mass, 
and a large enhancement 
due to a broad structure at low $\Kp\Km$ invariant mass. 
To further clarify these structures, we show in Fig.~\ref{fig:fXplots} 
the respective invariant mass projections following requirements 
that remove low mass combinations on the other axis of the Dalitz plane.
Approximately half of our signal events appear to originate
from the structure at low $\Kp\Km$ invariant mass.
We have studied the Dalitz plot distributions of the backgrounds,
which are found to be consistent with expectations,
and do not contain any structures that may explain the
peak in the $\Kp\Km$ invariant mass distribution.
Further interpretation of this structure and the rest of the 
$\Bp\to\Kp\Km\pip$ Dalitz plot requires an amplitude analysis.

In summary, we have made the first measurement of 
the charmless hadronic $B$ decay branching fraction 
${\cal B}(\Bu\to\Kp\Km\pip) = \kkpiBFwe$.
The \CP\ asymmetry is found to be consistent with zero.
Inspection of the Dalitz plot of signal candidates 
shows a broad structure peaking near
1.5\,\gevcc\ in the $\Kp\Km$ invariant mass distribution
that is reminiscent of similar structures seen in other 
charmless multibody hadronic $B$ decays~\cite{Garmash:2004wa,Aubert:2006nu,babar:kkks,Aubert:2005ce}.
This is likely to be of great interest for the understanding of 
low energy hadronic bound states~\cite{Klempt:2007cp}.

We are grateful for the excellent luminosity and machine conditions
provided by our \pep2\ colleagues, 
and for the substantial dedicated effort from
the computing organizations that support \babar.
The collaborating institutions wish to thank 
SLAC for its support and kind hospitality. 
This work is supported by
DOE
and NSF (USA),
NSERC (Canada),
CEA and
CNRS-IN2P3
(France),
BMBF and DFG
(Germany),
INFN (Italy),
FOM (The Netherlands),
NFR (Norway),
MES (Russia),
MEC (Spain), and
STFC (United Kingdom). 
Individuals have received support from the
Marie Curie EIF (European Union) and
the A.~P.~Sloan Foundation.


\begin{thebibliography}{99}
\bibitem{cc}
  Inclusion of the charge conjugate mode is implied
  except where explicitly stated otherwise.
\end{thebibliography}
\end{document}